\documentclass[prl,twocolumn,showpacs,preprintnumbers,amsmath,amssymb]{revtex4}
\usepackage{graphicx}
\begin{document}
\title{
LO Phonon-Induced Exciton Dephasing in Quantum Dots: An Exactly
Solvable Model}
\author{
E. A. Muljarov$^{1,2,}$}\email{muljarov@gpi.ru}
\author{R. Zimmermann$^1$}
\affiliation{ $^1$Institut f\"ur Physik der Humboldt-Universit\"at
zu Berlin,
Newtonstrasse 15, D-12489 Berlin, Germany\\
$^2$General Physics Institute, Russian Academy of Sciences,
Vavilova 38, Moscow 119991, Russia
 }
\begin{abstract}
It is widely believed that, due to its discrete nature, excitonic
states in a quantum dot coupled to dispersionless LO phonons form
everlasting mixed states (exciton polarons) showing no line
broadening in the spectrum. This is indeed true if the model is
restricted to a limited number of excitonic states in a quantum
dot. We show, however, that extending the model to a large number
of states results in LO phonon-induced spectral broadening and
complete decoherence of the optical response.
 \end{abstract}
\date\today

\maketitle

Among different mechanisms of the optical decoherence in
semiconductor quantum dots (QDs), interaction between excitons and
lattice vibrations (phonons) is the most important one at low
carrier densities, leading to a temperature-dependent linewidth in
optical spectra. The discrete nature of excitonic levels in a QD
gave rise to the so-called phonon bottleneck
problem~\cite{Bockelmann90}: When the phonon energy does not fit
the level separation, real phonon-assisted transitions between
different excitonic states are not allowed. On the other hand, the
measured optical polarization in QDs shows a partial initial
decoherence and a temperature-dependent exponential decay at
larger times~\cite{Borri01}. The bottleneck problem is partly
removed for acoustic phonons as they have an energy dispersion and
thus contribute to carrier relaxation~\cite{Bockelmann93}.
Moreover, even apart from the resonance between the phonon energy
and the level separation, acoustic phonons, due to their
dispersion, are shown to be responsible for pure dephasing induced
by virtual processes~\cite{Muljarov04}.

Longitudinal optical (LO) phonons, in turn, have a negligible
dispersion which leads to a quite different behavior: Excitons in
QDs and LO phonons are always in a strong coupling
regime~\cite{Hameau99} in the sense that they form everlasting
polarons with {\it no spectral
broadening}~\cite{Stauber00,Verzelen02,Jacak03,Stauber06}. This
also makes any approximate perturbative approach to the exciton-LO
phonon problem inappropriate. Indeed, for a few excitonic levels
in a QD coupled to bulk LO phonons, a self-consistent second Born
approximation for the self energy developed
in~\cite{Inoshita97,Kral98} results in a line broadening which is
fully artificial. This artefact is refuted by the exact solution
of this problem~\cite{Stauber00,Stauber06} showing that the
spectrum consists exclusively of discrete unbroadened lines. Also,
a Gauss spectral lineshape was found in the quadratic coupling
model by truncating the cumulant expansion in second
order~\cite{Uskov00}. Again, the exact solution of this model
shows no spectral broadening~\cite{Muljarov06}.

In this Letter we show that pronounced LO phonon-induced dephasing
and spectral broadening do nevertheless exist in QDs. This
broadening is calculated microscopically by inclusion of
infinitely many excitonic states in a QD that has never been done
before. Thus a qualitatively new source of the dephasing in QDs is
found.

The full problem of excitons in a QD linearly coupled to the LO
phonon displacement can be solved exactly only for a very limited
number of excitonic states~\cite{Stauber06}. The major obstacle
are off-diagonal terms in the exciton-phonon interaction which
couple different excitonic states in a QD. In order to take into
account as many excitonic states as we like, we have derived
microscopically an effective exciton-phonon Hamiltonian which has
only level-diagonal terms and thus allows an exact solution for an
arbitrary number of states. This effective Hamiltonian, however,
preserves the main features of the original problem, since the
off-diagonal terms are also intrinsically represented: By means of
a unitary transformation they are mapped into diagonal terms
giving rise to an exciton-phonon coupling quadratic in the phonon
displacement operators~\cite{Dean70,Muljarov04}.

Concentrating on the ground exciton state $|1\rangle$, the
effective Hamiltonian takes the form ($\hbar=1$):
\begin{eqnarray}
H&=& \omega_0 \sum_{\bf q}\, a^\dagger_{\bf q} a_{\bf
q} + (E_1+V_L+V_Q) |1\rangle \langle 1|,\label{H}\\
V_{L}&=& \sum_{\bf q} M_{11}({\bf q})\,( a_{\bf q} + a_{\bf
-q}^\dagger)\,,\label{V_L}\\
V_{Q}&=& -\frac{1}{2} \sum_{\bf p\, q} Q({\bf p},{\bf q}) \,(
a_{\bf p} +
a_{\bf -p}^\dagger ) ( a_{\bf q} + a_{\bf -q}^\dagger )\,,\label{V_Q}\\
Q({\bf p, q})&=& 2\sum_{n\neq
1}\frac{E_n-E_1}{(E_n-E_1)^2-\omega_0^2}\,M_{1n}({\bf p}) M_{n1}({\bf q})\nonumber\\
&\equiv&\sum_{n\neq1}  F_n({\bf p}) F_n({\bf q})\,, \label{Q}
\end{eqnarray}
where $\omega_0$ is the dispersionless LO phonon frequency and
$E_n$ is the bare transition energy of a single-exciton state
$|n\rangle$ in a QD. The kernel $Q$ of the quadratic coupling to
LO phonons has the same form as the effective scattering matrix
used in Ref.\,\cite{Dean70} to describe the phonon modes bound to
a neutral donor. It is derived from the level-nondiagonal matrix
elements \mbox{$M_{1n}({\bf q})\propto q^{-1}\langle 1|e^{i {\bf
qr}_e}-e^{i {\bf qr}_h}|n\rangle$} of the linear exciton-phonon
(Fr\"ohlich) interaction, up to second order. Using its
factorization property, $Q$ is expressed in Eq.\,(\ref{Q}) in
terms of functions $F_n({\bf q})$. A quadratic coupling model
similar to Eqs.\,(\ref{H}--\,\ref{Q}) has been already
successfully used for calculation of the exciton dephasing in
InGaAs QDs, induced by acoustic phonon-assisted virtual
transitions~\cite{Muljarov04}.

The method developed in~\cite{Muljarov04} allows us to find the
{\it exact solution} of the Hamiltonian~(\ref{H}--\,\ref{Q}),
using the cumulant expansion. The linear optical polarization has
the form
\begin{equation}
P(t)=\theta(t)\exp[-iE_1\,t+K_L(t)+K_Q(t)+K_M(t)]\,, \label{P}
\end{equation}
where the linear and quadratic cumulants are
\begin{eqnarray}
K_L(t)&=&\left[(2N+1)(\cos\omega_0t-1)+i(\omega_0t-\sin\omega_0t)\right] \nonumber\\
&&\times \sum_{\bf q} |M_{11}({\bf q})/\omega_0|^2, \label{K_L}\\
K_Q(t)&=&-\frac{1}{2}\sum_{\nu\,j}
\ln[1-i\Lambda_\nu\lambda_j(t)]\,.\label{K_Q}
\end{eqnarray}
Here $\Lambda_\nu$ are the eigenvalues of the matrix
\begin{equation}
A_{nm}=\sum_{\bf q} F_n(-{\bf q}) F_m({\bf q})\,, \label{A}
\end{equation}
while $\lambda_j(t)$ are the eigenvalues of the Fredholm problem
\begin{equation}
\int_0^t d\tau' D(\tau-\tau')
u_j(\tau';t)=\lambda_j(t)u_j(\tau;t)\,. \label{Fredholm}
\end{equation}
Equation~(\ref{Fredholm}) can be solved numerically, as it has
been done for acoustic phonons~\cite{Muljarov04}. However, for
dispersionless optical phonons the propagator $D$ is ${\bf
q}$-independent:
$D(\tau)=(N+1)e^{-i\omega_0|\tau|}+Ne^{i\omega_0|\tau|}$ [where
$N=1/(e^{\beta\omega_0}-1)$ and $\beta=1/k_BT$] and thus allows an
analytic solution of Eq.\,(\ref{Fredholm}) which can be found in
Ref.~\cite{Muljarov06}. Finally, the mixed cumulant $K_M(t)$ in
Eq.\,(\ref{P}) is found in the same way as the quadratic one but
additionally requires the eigenvectors of the matrix $A_{nm}$ and
the eigenfunctions $u_j$ of Eq.\,(\ref{Fredholm}). It turns out,
however, that it leads to only tiny corrections:
$|K_M/K_Q|<10^{-7}$ in CdSe QDs considered here, and thus can be
safely neglected.

We have already used the model Eqs.\,(\ref{H}--\,\ref{Q}) and its
solution Eqs.\,(\ref{P}\,--\ref{Fredholm}) in case of two
excitonic levels in an InGaAs QD~\cite{Muljarov06} to show that a
truncation of an infinite series in the cumulant as done in
Ref.~\cite{Uskov00} leads to an artificial Gauss decay of the
optical polarization. Only taking into account an infinite number
of all-order diagrams gives the correct result which is an almost
perfectly periodic time-dependent polarization. In the present
Letter, using the same model, we include now (infinitely) many
excitonic states in a QD and show that this leads to a {\it
qualitatively new effect:} polarization decay and spectral
broadening.

This effect takes place in any type of semiconductor QDs. For the
sake of illustration, we concentrate on the polar material CdSe
and a simple model of a QD. Material parameters (Fr\"ohlich
coupling constants $\alpha_e=0.47$, $\alpha_h=0.88$;
$\hbar\omega_0=24$\,meV) and the QD model are the same as in
Ref.~\cite{Stauber06}, but the parabolic confinement potentials
are taken as isotropic. For the Gauss localization length of
electron and hole we use here three sets of parameters: (A)
$l_e=l_h=2.5$\,nm; (B) $l_e=2.71$\,nm, $l_h=2.2$\,nm (as in
Ref.~\cite{Stauber06}); and (C) $l_e=4.14$\,nm, $l_h=3.36$\,nm,
which gives half the level distance compared to case (B). We
decided to neglect the Coulomb interaction as it would only
renormalize the transition energies $E_n$ and induce minor changes
in the polarization dynamics of the ground exciton state.

\begin{figure}[b]
\includegraphics*[angle=-90,width=6.5cm]{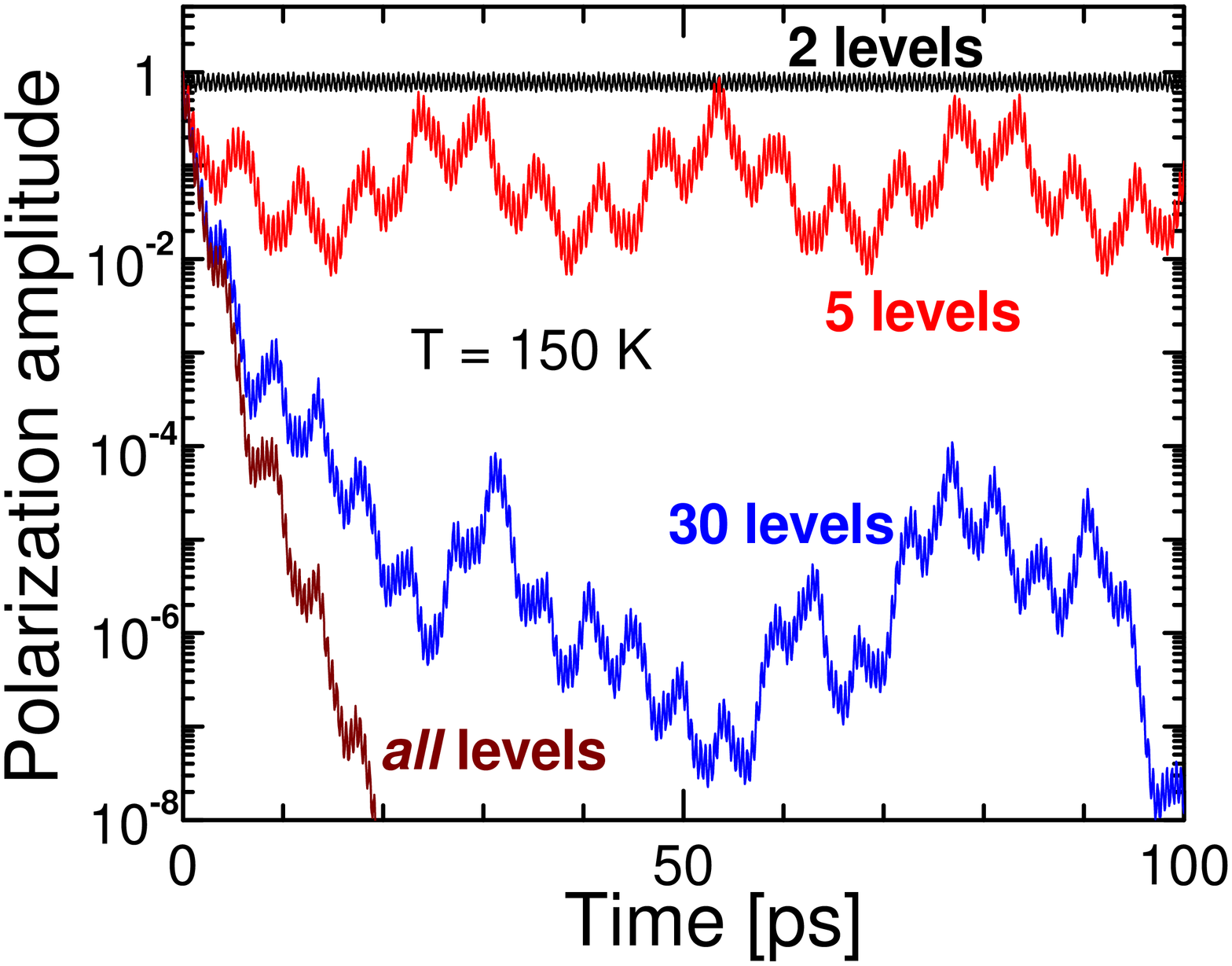}%
\caption{Amplitude of the linear optical polarization of the
ground exciton state at $T=150$\,K, taking into account two, five,
thirty, and all exciton levels in a CdSe QD (set A).}
\end{figure}
The amplitude of the time-dependent linear polarization after
impulsive excitation in a CdSe QD at $T=150$\,K is shown in
Fig.\,1 for set (A). In this case, due to the full charge
neutrality of the electron-hole pair, $M_{11}({\bf q})=0$ holds,
and the linear term $V_L$ vanishes. Thus we concentrate at the
moment on the quadratic term $V_Q$, Eq.\,(\ref{V_Q}), which is in
fact the only source of dephasing. For two excitonic levels [i.e.,
only one excited state $n=2$ contributes to Eq.\,(\ref{Q}), and
the matrix $A_{nm}$, Eq.\,(\ref{A}), reduces to a scalar] the
polarization has only tiny oscillations close to unity. For five
levels, it decays and then revives almost to unity after 20\,ps.
For 30 levels, the amplitude decays already dramatically and then
oscillates irregularly around a small value. Such a behavior has
been already discussed before and called
quasi-dephasing~\cite{Stauber06,Jacak03a}. Finally, if we take
into account all excitonic states, the polarization decays
strictly: At 100\,ps its amplitude drops to $10^{-30}$ (but is
shown in Fig.\,1 only up to $10^{-8}$). In practice, of course, we
truncate the infinite matrix $A_{nm}$ and check that a larger
matrix produces no changes within the time interval considered.

To understand better why the optical polarization becomes decaying
when more and more exciton (electron-hole pair) states are taken
into account, we calculate the linear absorption spectrum, i.e.
the real part of the Fourier transform of $P(t)$. It is hopeless,
however, to find the Fourier transform numerically if the function
does not decay at all or decays very weakly. That is why we use
here a different method: We diagonalize exactly the excited state
Hamiltonian $H_1$ which is related to the full one,
Eq.\,(\ref{H}), as $H=H_0|0\rangle\langle 0| + H_1|1\rangle\langle
1|$. Such a diagonalization is very easy if $V_L=0$. In case of
{\it two} levels $H_1$ becomes
\begin{equation}
H^{\rm 2lev}_1 = \Omega B^\dagger B + \tilde{E_1} + \omega_0
\sum_\mu b^\dagger_\mu b_\mu\,, \label{diag}
\end{equation}
where
\begin{eqnarray}
\Omega&=&\sqrt{\omega_0^2-2\omega_0\sum_{\bf q} F_2(-{\bf q})
F_2({\bf
q})}, \\
B&=&\sum_{\bf q} F_2({\bf q}) (\xi_+ a_{\bf q}+\xi_-
a^\dagger_{-\bf q})
\end{eqnarray}
are, respectively, the new phonon frequency and annihilation
operator,
$\xi_\pm=(\Omega\pm\omega_0)/2\sqrt{\omega_0\Omega\sum_{\bf q}
|F_2({\bf q})|^2}$, and $\tilde{E_1}$ is the polaron shifted
transition energy.

Note that out of a continuum of degenerate LO modes only a single
phonon mode couples to the QD and produces a new, {\it bound}
mode~\cite{Dean70} $B$ with frequency $\Omega$, while all the
other modes $b_\mu$, which show up in the last term of
Eq.\,(\ref{diag}), are decoupled. They are simply some orthogonal
linear combinations of the former phonon modes $a_{\bf q}$ and
have the same old frequency $\omega_0$. The linear polarization
$P(t)=\theta(t)\langle e^{iH_0t} e^{-iH_1t}\rangle$, calculated in
the same way as in Ref.~\cite{Stauber06}, then takes the form:
\begin{equation}
P^{\rm 2lev}(t)=\theta(t)\sum_{\mathbf{n,m}=0}^\infty
\alpha_{\mathbf{nm}}\exp\{i(\mathbf{n}\omega_0-\mathbf{m}\Omega)t\},
\label{P2lev}
\end{equation}
where $\alpha_{\mathbf{nm}}=(1-e^{-\beta\omega_0})e^{-\beta
\mathbf{n} \omega_0}|\mathbf{\langle n|m)}|^2$, and
$\mathbf{\langle n|m)}$ are the projections of new phonon states
$\mathbf{|m)}$ into old ones $\mathbf{|n\rangle}$ (to be
distinguished from exciton states $|n\rangle$), which are
calculated recursively.

Generalization to ${\cal N}$ excitonic levels is straightforward:
$H_1$ is now diagonalized to ${\cal N}-1$ new phonon modes:
\begin{equation}
\Omega_\nu=\sqrt{\omega_0^2-2\omega_0\Lambda_\nu}
\end{equation}
with $\Lambda_\nu$ being the eigenvalues of the $({\cal
N}-1)$-dimensional matrix Eq.\,(\ref{A}). In accordance with
Eqs.\,(\ref{P}) and (\ref{K_Q}), the full linear polarization can
be written as a product
\begin{equation}
P(t)=\prod_\nu P_\nu(t) \label{Pfull}
\end{equation}
of functions $P_\nu(t)$ due to each individual phonon mode given
by the same Eq.\,(\ref{P2lev}), where $\Omega$ and
$\alpha_{\mathbf{nm}}$ are replaced, respectively, by $\Omega_\nu$
and $\alpha^\nu_{\mathbf{nm}}$.

\begin{figure}[t]
\includegraphics*[angle=0,width=6.5cm]{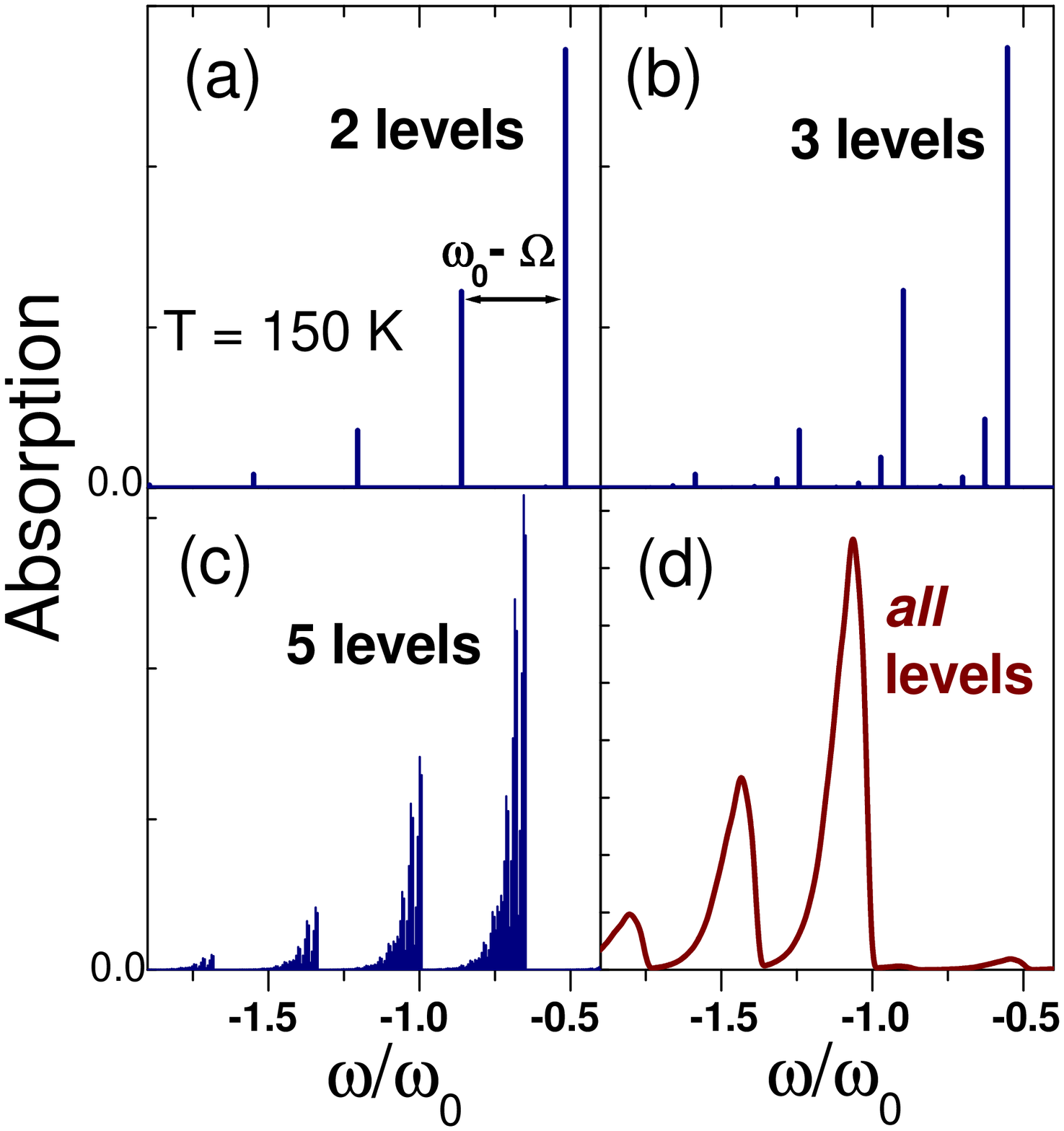}%
\caption{Evolution of the fine structure in the absorption
spectrum of a QD by accounting for more and more excitonic levels
[linear scale, set (A), $E_1$ is taken as zero of energy].}
\end{figure}

For two levels, the absorption spectrum, already discussed in
Ref.~\cite{Muljarov06}, represents a set of discrete lines. There
are two-phonon satellites around the zero-phonon line (the
standard one-phonon satellites are absent here as $V_L=0$), but
more important is the fine structure shown in Fig.\,2\,(a). This
fine structure is due to the difference between the old and the
new phonon frequency $\omega_0-\Omega$ and manifests itself in the
time domain as rapid oscillations with the period of 0.5\,ps
(Fig.\,1). Including the third excitonic level brings in an
additional frequency and more lines in the spectrum, Fig.\,2\,(b).
With five levels (and four new phonon frequencies), there is
already a plenty of closely lying delta lines, Fig.\,2\,(c).
Finally, if we include all exciton levels, these lines merge and
produce a {\it continuous broadening}, Fig.\,2\,(d). In the time
domain, the polarization represents a superposition of individual
oscillations [see Eqs.\,(\ref{Pfull}) and (\ref{P2lev})] which
interfere destructively due to incommensurability of their
frequencies and thus lead to dephasing.

\begin{figure}[t]
\includegraphics*[angle=-90,width=7.9cm]{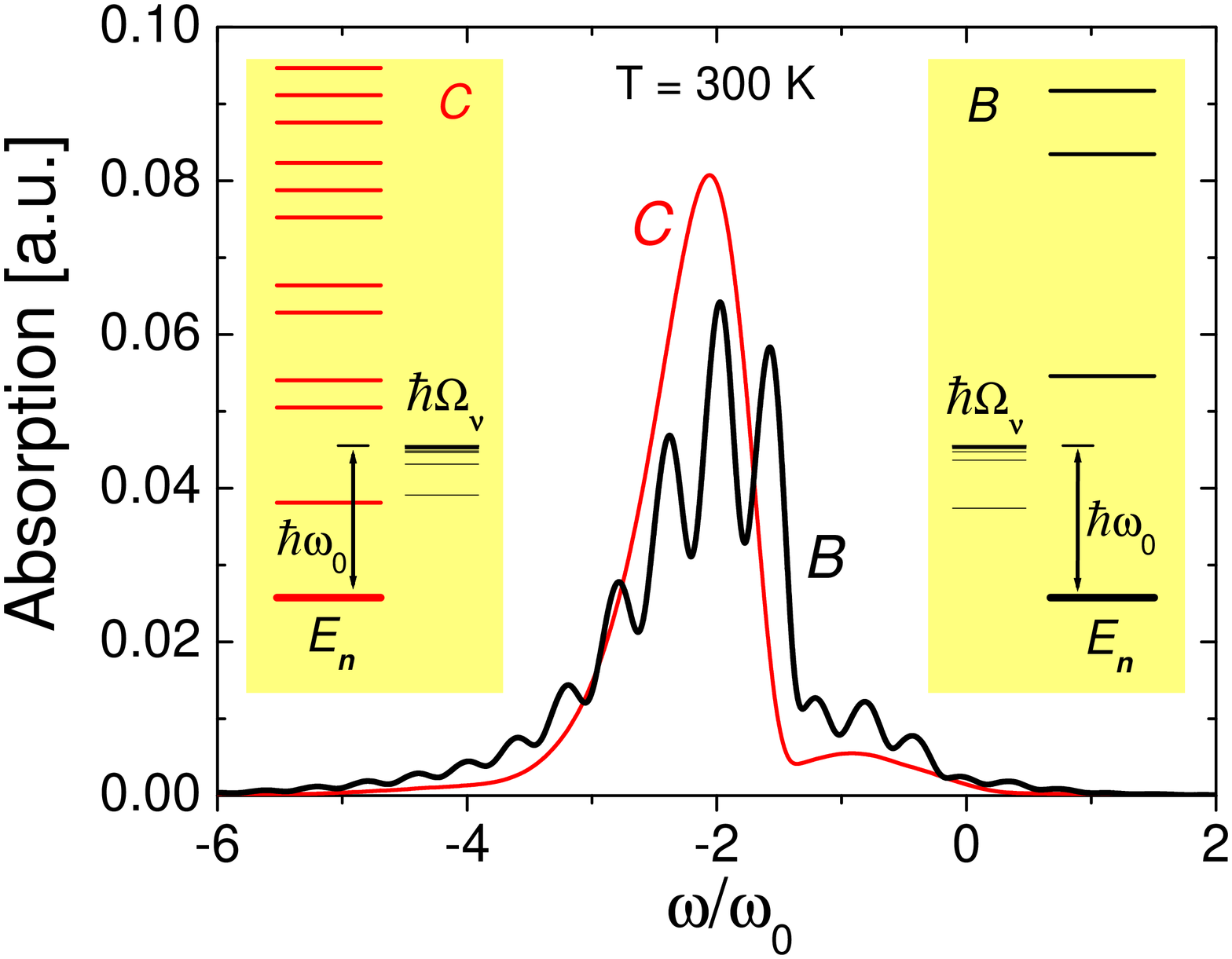}%
\caption{Absorption spectrum  of a CdSe QD at $T=300$\,K,
calculated for two different types of the bare-exciton level
structure [sets (B) and (C), see text]. Bound phonon frequencies
$\Omega_\nu$ and levels $E_n$ of exciton excited (ground) states
are shown in the insets by thin (thick) horizontal lines.}
\end{figure}

In contrast to the present model of a QD with discrete exciton
levels only, in realistic QD systems there is also a wetting-layer
continuum. This continuum could seem to be a more probable
candidate for producing the spectral broadening~\cite{Seebeck05}.
It turns out, however, that in typical QDs, there are usually
enough discrete exciton levels to provide an appreciable
polarization decay. In fact, taking into account 30 levels is
already sufficient to have the dephasing time very close to the
exact one (i.e., with all levels, see Fig.\,1), but at the same
time it requires the heterostructure potential band offset to be
less than 1.0\,eV for the electron and 0.5\,eV for the
hole~\cite{footnote}. We account for all discrete levels in the
harmonic potentials just for consistency, in order to show that a
complete decay of the polarization is indeed achieved.

The full absorption spectrum of a CdSe QD is shown in Fig.\,3. As
we have in sets (B) and (C) different Gauss lengths, $l_e\neq
l_h$, the linear coupling $V_L$ contributes too, which gives
surprisingly only a slight modification in the spectrum. For set
(B) all the excited levels are well above $\hbar\omega_0$. All the
bound-phonon frequencies $\Omega_\nu$ are condensed just below
$\omega_0$, except one which is well separated and is in fact
responsible for the spectral fine structure. This fine structure
is absent if we take a larger dot with closer levels (Fig.\,3, red
curve and inset C). Here, one excited level is found below
$\hbar\omega_0$, and even two bound phonons are well resolved and
wash out the fine
\begin{figure}[b]
\includegraphics*[angle=-90,width=7.5cm]{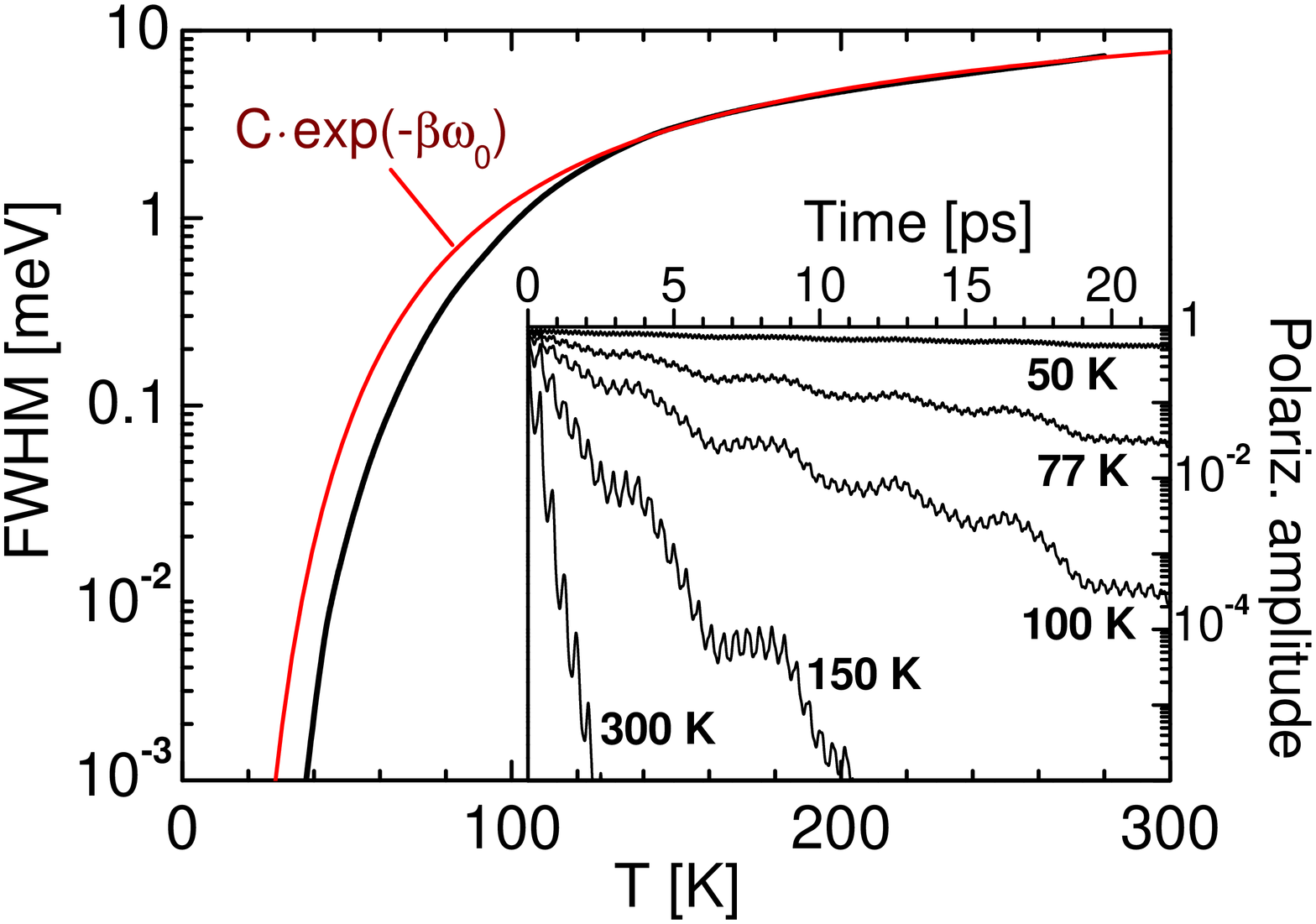}%
\caption{Spectral linewidth of the ground exciton state in a CdSe
QD, set (B), as a function of temperature (solid curve) and the
exponential fit with $C=20$\,meV (dashed curve). Inset:
Polarization amplitude at different temperatures.}
\end{figure}
structure. Thus, changes in the exciton level structure modify the
spectrum but obviously do not affect our principal conclusion on
the finite spectral width, and even do not alter much the
linewidth itself.

Since the decay is not strictly exponential (see the inset in
Fig.\,4), we have extracted the linewidth directly from the
absorption and plotted in Fig.\,4 the full width at half maximum
(FWHM) as a function of temperature. At high temperatures, the
FWHM varies from few to ten meV and can be well approximated by an
exponential law with $\hbar\omega_0$ being the activation energy.
Below 50\,K the linewidth is practically unaffected by LO phonons,
and the dephasing in QDs is determined by some other mechanisms
like coupling to acoustic phonons~\cite{Muljarov04}, phonon
anharmonicity~\cite{Machnikowski06}, and radiative decay.

In conclusion, we have calculated the time-dependent optical
polarization and the absorption spectrum of a quantum dot using
the exactly solvable model of excitons quadratically coupled to LO
phonons. We have extended this model to an arbitrary number of
excitonic states in a quantum dot and (using its exact solution)
demonstrated that taking into account infinitely many states
results in an LO phonon-induced exciton dephasing and spectral
broadening. \smallskip

Financial support by DFG Sonderforschungsbereich 296 is gratefully
acknowledged. E.\,A.\,M. acknowledges partial support by Russian
Foundation for Basic Research and Russian Academy of Sciences.


\begin{thebibliography}{10}
\bibitem{footnote} Electron (hole) energy levels are given by
$(2n+l)\Delta_{e(h)}$, where $\Delta_e=80$\,meV and
$\Delta_h=35$\,meV~\cite{Stauber06} in case (B); 30 levels are
produced by taking $0\leqslant n \leqslant 4$ and $0\leqslant l
\leqslant 5$.
\end{thebibliography}
\end{document}